\documentclass[Letter]{aa}
\usepackage[varg]{txfonts}
\usepackage{booktabs,graphicx,siunitx,xspace,natbib}
\usepackage{subfig}
\usepackage{ragged2e} 
\usepackage{array}
\DeclareSIUnit \parsec {pc}
\DeclareSIUnit \pc     {\parsec}
\DeclareSIUnit \kpc    {\kilo \parsec}
\DeclareSIUnit \msun   {\ensuremath{M_\sun}}
\DeclareSIUnit \year   {a}

\newcommand{\Msol}{M\ensuremath{_\odot}\xspace}

\newcommand{\about}{$\sim$}

\newcommand{\pos}{$\mathrm{e^+}$}
\newcommand{\ele}{$\mathrm{e^-}$}

\begin{document}

\title{Search for 511~keV Emission in Satellite Galaxies of the Milky Way with INTEGRAL/SPI}

\author{
  Thomas Siegert   					  \inst{\ref{inst:mpe}}\thanks{E-mail: tsiegert@mpe.mpg.de} \and
  Roland Diehl    				    \inst{\ref{inst:mpe},\ref{inst:xcu}} \and
  Aaron C. Vincent  				  \inst{\ref{inst:ippp}} \and
	Fabrizia Guglielmetti       \inst{\ref{inst:mpe},\ref{inst:mpa}}  \and
	Martin G. H. Krause					\inst{\ref{inst:sps}} \and
  Celine Boehm     				    \inst{\ref{inst:ippp}} 
}
\institute{
  Max-Planck-Institut f\"ur extraterrestrische Physik, Gie\ss enbachstra\ss e, D-85741 Garching, Germany
  \label{inst:mpe}
 \and
  Excellence Cluster Universe, Boltzmannstra\ss e 2, D-85748, Garching, Germany
  \label{inst:xcu}
 \and
	Institute for Particle Physics Phenomenology, Department of Physics, Durham University, Durham DH1 3LE, United Kingdom
	\label{inst:ippp}
 \and
  Max-Planck-Institut f\"ur Astrophysik, Karl-Schwarzschild-Stra\ss e 1, D-85748 Garching, Germany
  \label{inst:mpa}
 \and
	School of Physical Sciences, University of Tasmania, Hobart, TAS, 7005, Australia
	\label{inst:sps}
  }
\date{Received 17 Jun 2016 / Accepted 30 Jul 2016}

\abstract
{The positron (\pos) annihilation $\gamma$-ray signal in the Milky Way (MW) shows a puzzling morphology: a very bright bulge and a very low surface-brightness disk. A coherent explanation of the \pos origin, propagation through the Galaxy and subsequent annihilation in the interstellar medium has not yet been found. Tentative explanations involve \pos s from radioactivity, X-ray binaries, and dark matter (DM).}
{Dwarf satellite galaxies (DSGs) are believed to be DM-dominated and hence are promising candidates in the search for 511 keV emission as a result of DM annihilation into \pos\ele-pairs. The goal of this study is to constrain possible 511 keV $\gamma$-ray signals from 39 DSGs of the MW and to test the annihilating DM scenario.}
{We use the spectrometer SPI on INTEGRAL to extract individual spectra for the studied objects in the range 490--530 keV. As the diffuse galactic 511 keV emission dominates the overall signal, the large scale morphology of the MW has been modelled accordingly and was included in a maximum likelihood analysis. Alternatively, a distance-weighted stacked spectrum has been determined, representing an average DSG seen in 511 keV.}
{Only Reticulum II (Ret II) shows a $3.1\sigma$ signal. Five other sources show tentative $2\sigma$ signals. The mass-to-511~keV-luminosity-ratio, $\Upsilon_{511}$, shows a marginal trend towards higher values for intrinsically brighter objects, opposite to the mass-to-light-ratio, $\Upsilon_V$ in the V-band, which is generally used to uncover DM in DSGs.}
{All derived 511~keV flux values or upper limits are above the flux level implied by a DM interpretation of the MW bulge signal. The signal detected from Ret II is unlikely to be related to a DM origin alone, otherwise, the MW bulge would be $\sim100$ times brighter in 511~keV than what is seen with SPI. Ret II is exceptional considering the DSG sample, and rather points to enhanced recent star formation activity, if its origins are similar to processes in the MW. Understanding this emission may provide further clues regarding the origin of the annihilation emission in the MW bulge.}

\keywords{
  Positrons,
  Gamma rays: general,
	ISM: general,
	Galaxies: dwarf satellites,
  Techniques: spectroscopic,
	Cosmology: dark matter
}

\maketitle

%
\section{Introduction}
\label{sec:intro}
It has been proposed that the 511~keV morphology of the Milky Way (MW), originating in the annihilation of electrons (\ele s) with positrons (\pos s), seen by INTEGRAL \citep{Winkler2003_INTEGRAL}, could be related to the decay or annihilation of dark matter (DM) particles \citep{Hooper2004_dm,Ascasibar2006_511dm}. From theoretical considerations it was suggested that when light DM particles ($1~\mathrm{MeV~c^{-2}}\lesssim m_{\chi} \lesssim100~\mathrm{MeV~c^{-2}}$) annihilate or decay, they could produce \pos s with low kinetic energies of \about$\mathrm{MeV}$~\citep{Boehm2004_dm,Hooper2004_dm,Picciotto2005_dm,Beacom2006_511,Gunion2006_dm,Pospelov2008_dm,Boehm2008_dm}. The annihilation of these \pos s with \ele s from the interstellar medium (ISM) would lead to the signature that was measured by the spectrometer SPI \citep{Vedrenne2003_SPI} on INTEGRAL.

The galactic diffuse large-scale 511~keV emission that was measured with balloon-flight experiments \citep[e.g.][]{Leventhal1978_511} and with SPI \citep{Knoedlseder2005_511,Bouchet2010_511,Skinner2014_511} was found to be concentrated towards the bulge region of the MW, reminiscent of a DM halo profile. However, other -- less exotic -- sources may also explain this signal \citep[see][for a review]{Prantzos2011_511}.

If the entire bulge annihilation radiation originates from DM particles, the apparently DM-dominated dwarf satellite galaxies (DSGs) of the MW should also emit a measurable 511~keV signal~\citep{Hooper2004_dm,Simon2007_dm,Strigari2008_dm}. Based on cold dark matter (CDM) cosmology and the corresponding galaxy formation model \citep[see e.g.][]{White1978_galaxies,Springel2005_milsim,Moster2013_galaxies}, the satellite galaxies of the MW must be DM-dominated \citep{Mateo1998_dsphs,Strigari2008_dm_dsph,McConnachie2012_dsph}.

A good test of the annihilating DM hypothesis is thus to check in cumulative INTEGRAL data for a consistent 511~keV brightness from the known satellites of the MW, depending on their DM content and distance. \citet{Cordier2004_511dm} tested this for the case of the Sagittarius Dwarf Spheroidal (Sag). A point-like emission, as expected from DM annihilation (see below), could not be detected, and a $2\sigma$ upper limit on the 511~keV flux of $2.5 \times 10^{-4}~\mathrm{ph~cm^{-2}~s^{-1}}$ was established. They could neither exclude nor corroborate DM as the cause of the 511~keV emission in the MW because the upper limit from Sag compared to the MW bulge flux was not constraining enough. However, this is based on the assumption that the whole 511~keV emission in the bulge of the MW arises from annihilating DM.

In this work, we extend and refine these previous studies, using more than ten years of INTEGRAL/SPI data covering the full sky. We report a new search for point-like 511~keV line emission at the positions of 39 DSGs of the MW within 500 kpc in order to provide new constraints on a DM origin of the galactic positron signal. We also report on the discovery of a tentative signal in the Reticulum II dwarf galaxy.

\begin{figure}
  \centering
  \includegraphics[width=\linewidth]{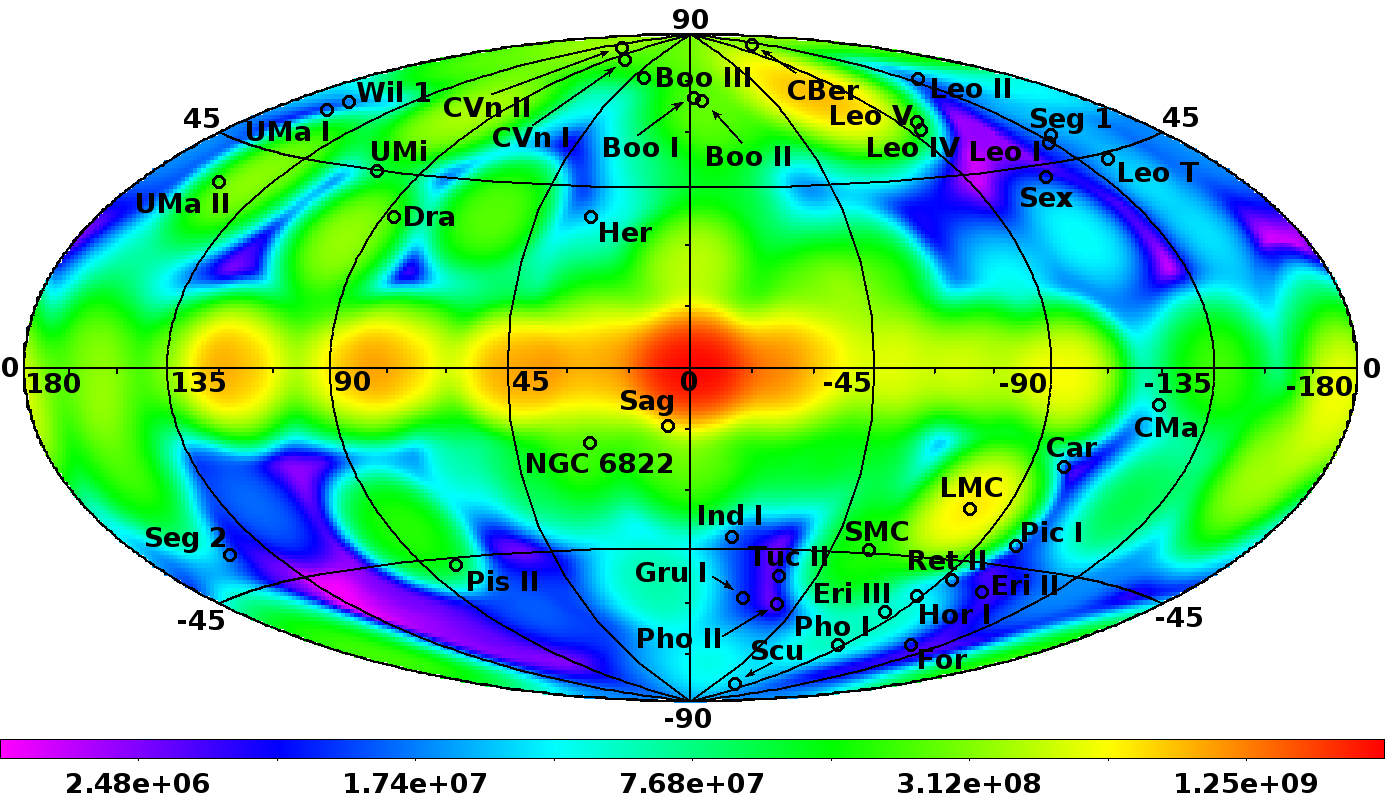}
   \caption{Sky exposure with SPI after 1258 orbits of the INTEGRAL mission. Tested satellite galaxies of the Milky Way are marked with a circle. Colourbar shows the SPI exposure in units of $\mathrm{cm^2~s}$. The effective area of SPI at photons energies of 511~keV is $\approx75~\mathrm{cm^2}$.}
  \label{fig:SPI-exposure}
\end{figure}

%
\section{Data and their Analysis}
\label{sec:spi-analysis}
%
\subsection{Data Set, Background Model, and Celestial Large-Scale Emission}

The data that we analysed in this work were taken between 12 Dec 2002 and 7 Apr 2013 with the spectrometer SPI on ESA's INTEGRAL satellite, and are identical to the data set of \citet[][hereafter Paper I]{Siegert2016_511}. Therefore, we refer to this paper for detailed information on the data selection and analysis procedure.

In total, 73590 pointings with an overall exposure time of 160~Ms were analysed. We show the exposure map in Fig.~\ref{fig:SPI-exposure}, together with the positions of the 39 investigated DSGs. We used the maximum likelihood method to compare the measured data to models of celestial emission  and background. In particular, the modelled time-patterns for each model component (see below) are fitted to the measured time-pattern of the data by maximising the likelihood of the data, i.e. estimating intensity scaling parameters for each sky and background component individually. In this maximum likelihood method, we account for photon count statistics being Poisson-distributed, and use the Cash statistic \citep{Cash1979_cstat} for measured data $d_k$ and modelled data $m_k = \sum_{i} \theta_i M_{ik}$ with model components $M_{ik}$ for instrumental backgrounds and celestial signals (see Paper I for more details):
\begin{equation}
C(D|\theta_i) = 2 \sum_{k} \left[ m_k - d_k \ln m_k \right],
\label{eq:cstat}
\end{equation}
where $\theta_i$ are the individual intensity scaling parameters. By using the Cash statistic, the corresponding model $m_k$ is positive definite in any case, avoiding an issue of negativity in data $d_k$ that may occur if simple background subtraction would be applied instead. The goodness of fit for the baseline model was shown to be sufficient in Paper I.

Instrumental background is modelled by a self-consistent description of the data, treating instrumental line and continuum backgrounds separately (see Paper I). The focus of this work is to search for additional point-like 511~keV \pos annihilation $\gamma$-ray signals beyond the diffuse large-scale emission. The overall emission from annihilating \pos s in the MW dominates the signal and hence a large scale emission model is needed to avoid possible falsely attributed emission. We adopt the large-scale emission model of Paper I which describes the Galaxy in 511~keV well through an empirical six-component model, made of 2D-Gaussians with different positions and sizes mapped onto the sky. In particular, this model consists of three extended components, which describe the inner Galaxy (narrow and broad bulge, respectively), and a thick low surface-brightness disk. Additionally, in the centre of the Galaxy, a point-like source is included \citep{Skinner2014_511,Siegert2016_511}. The two strongest continuum sources, the Crab and Cygnus X-1, are additional components of this model. We use these model components as a baseline model and superimpose the additional 39 DSGs, modelled as point-sources (see below), at their visible positions in the sky, while still allowing each baseline model components to vary in intensity, independently. 

The background scaling parameters and the six celestial scaling parameters from Paper I were re-determined in the maximum likelihood parameter estimation to account for possible enhanced contributions from the DSGs.

\subsection{Emission from Satellite Galaxies}

The focus of our study is to search for 511~keV gamma-ray line signals which are produced when \pos s find \ele s to annihilate with\footnote{When \pos s find \ele s, the resulting spectrum depends on the kinetic energies of the particles. Below a threshold of 6.8~eV, \ele s and \pos s can form an intermediate bound state, the positronium atom, which decays to either two 511~keV photons, or three continuum photons, distributed between 0 and 511~keV, see \citet{Ore1949_511}.}, either free or bound in atoms. The emissivity of 511~keV photons produced per unit time is driven by the annihilation conditions in the MW or in DSGs. These conditions include the number densities of \pos s and H-atoms as well as the ionisation fraction of H in the galactic ISM. While these number densities are large in the MW, there is no observational indication yet that there are similar large number densities in DSGs. This is unsurprising, as most of these objects contain only a small number of stars. In what follows, we nonetheless assume the number densities of H-atoms, or free electrons in the ISM of each dwarf galaxy to be large enough for each \pos-population to efficiently annihilate and produce a 511~keV line.

For simplicity, we assume the annihilation signals from DSGs to be point-like. If the signal is indeed from DM annihilation, the annihilation rate is proportional to the integral of the DM density squared over the line-of-sight (J-factor)
\begin{equation}
J \equiv \int_{\Delta \Omega} d \Omega \int_{l.o.s.} \rho^2 d \ell,
\label{eq:jfactor}
\end{equation}
where the first integral is over the solid angle of the region of interest, and the second is over the line of sight, characterising the distribution of annihilating DM in an astrophysical system. Typical dark matter density profiles follow a power law in the inner regions, $\rho(r) \propto r^{-\gamma}$ \citep{Burkert1995_dm,Navarro1996_dm,Merritt2006_dm} with $0 < \gamma \lesssim 2$. The $\rho^2$ dependence of the J-factor thus yields a very sharply peaked signal, in most cases. Generally, compilations of dwarf galaxy J-factors in the literature \citep[e.g.][]{Ackermann2014_FermiLAT_dsphs,Evans2016_Jfactors} yield regions of interest that are smaller than the imaging resolution of SPI $\sim 2.7^\circ$, so that the point-like assumption is adequate. For example, the imaging resolution of SPI encompasses a physical region of more than 400~pc for the closest DSG in our sample, Canis Major (CMa), at a distance of 9~kpc.

Our input catalogue of all\footnote{During the write-up of this study, more DSGs have been found but have not been included in the analysis.} DSGs near the MW within 500~kpc holds 39 individual candidate sources. We use the baryonic centres of the DSGs as the positions of the point-sources, see Tab.~\ref{tab:dwarf_fits}. This leads to 39 additional intensity scaling parameters $\theta_i$ in the model fit to the observed data. These sources are at least separated by more than the imaging resolution of SPI ($2.7^{\circ}$), and thus the correlation between them (source confusion) is usually negligible. Exceptionally "close pairs" (see Fig.~\ref{fig:SPI-exposure}) are CVn I -- CVn II ($6.5^{\circ}$), Leo I -- Seg 1 ($3.8^{\circ}$), Leo IV -- Leo V ($2.8^{\circ}$), and Boo I -- Boo II ($1.7^{\circ}$), so that the flux values derived from the latter pair only should be considered with caution.

For each galaxy, an individual spectrum in the range 490--530~keV was extracted. Then, in each spectrum, we determined the flux of annihilation emission separately. Due to the individually low signals, we additionally consider an alternative \textit{stacking} approach for a DM hypothesis test. In this case, instead of deriving 39 individual spectra, we fix their relative fluxes according to their distances, assuming the same mass for all DSGs \citep{Strigari2008_dm_dsph}. This obtains a spectrum for a reference DSG at a chosen distance of $D_0 = 100~\mathrm{kpc}$. The resulting spectrum, however, would be dominated by the closest galaxy as the flux is proportional to the inverse distance squared, and may also be confused by the diffuse emission in the galactic plane and bulge, due to their partial correlation in the maximum likelihood approach. We try to avoid such a bias -- in the stacking procedure only -- by ignoring DSGs towards the galactic plane (between $|b|<10^{\circ}$), and galaxies closer than 25~kpc. Formally, the additional (now seventh, see Paper I) sky component is described by Eq.~(\ref{eq:stack})
\begin{equation}
F = \frac{\left<L_0\right>}{4 \pi D_0^2} \sum_{i=1}^{39} \delta(l-l_i) \delta(b-b_i) \left(\frac{D_0}{D_i}\right)^2\mathrm{.}
\label{eq:stack}
\end{equation}
Here $\left<L_0\right>$ is the (fitted) intrinsic mean luminosity for a basic DSG at a canonical distance of $D_0 = 100~\mathrm{kpc}$, corresponding to 39 individual sources, at positions $(l_i / b_i)$ in the sky, scaled by their distances $D_i$.

%
\section{Results}
\label{sec:results}

\subsection{Individual Sources}

We first validate the emission attributed to the diffuse large-scale 511~keV emission to obtain a robust reference model with respect to possible additional sources. We find the bright bulge and faint disk, as well as the Galactic Centre Source (GCS), the Crab and Cygnus X-1 with fluxes consistent with the results reported in Paper I. The flux values for bulge, disk and GCS are $(9.5\pm0.7)\times10^{-4}~\mathrm{ph~cm^{-2}~s^{-1}}$, $(16.7\pm3.6)\times10^{-4}~\mathrm{ph~cm^{-2}~s^{-1}}$, and $(0.8\pm0.2)\times10^{-4}~\mathrm{ph~cm^{-2}~s^{-1}}$, respectively. Continuum fluxes in the analysed 40~keV band are $(2.20\pm0.07)\times10^{-5}~\mathrm{ph~cm^{-2}~s^{-1}~keV^{-1}}$ for the Crab, and $(0.65\pm0.05)\times10^{-5}~\mathrm{ph~cm^{-2}~s^{-1}~keV^{-1}}$ for Cygnus X-1, also consistent with literature values \citep[see e.g.][]{Jourdain2009_Crab,Jourdain2012_CygX1}.

The derived spectra for each DSG near 511~keV were fitted by a Gaussian-shaped line with width fixed at 2.15~keV (instrumental resolution, FWHM) on top of a constant offset. The centroid was allowed to vary in the range 508--514~keV, corresponding to bulk motions of $|v_{Bulk}| \approx 1750~\mathrm{km~s^{-1}}$, to account for intrinsic movement of the satellites and statistical fluctuations. For non-positive results, a $2\sigma$ flux limit is estimated, for a line at 511~keV.

The strongest DSG signal that we find is from the position of Reticulum II (Ret II), with $3.1\sigma$ significance. Its line flux is $(17.0\pm5.4)\times10^{-5}~\mathrm{ph~cm^{-2}~s^{-1}}$. However, we caution that Ret II 511~keV emission may be too intense a signal to be interpreted as due to DM alone (see Discussion below). For the position of Sag, a 511~keV line significance of $2.3\sigma$ is found. Formally, the line flux is $(2.2\pm1.0)\times10^{-5}~\mathrm{ph~cm^{-2}~s^{-1}}$, consistent with the upper limits derived from \citet{Cordier2004_511dm}, with a now $\sim$100 times larger exposure at this position.

\begin{figure}[!ht]
  \centering
   \includegraphics[width=\linewidth,trim=2cm 2cm 2cm 2cm,clip]{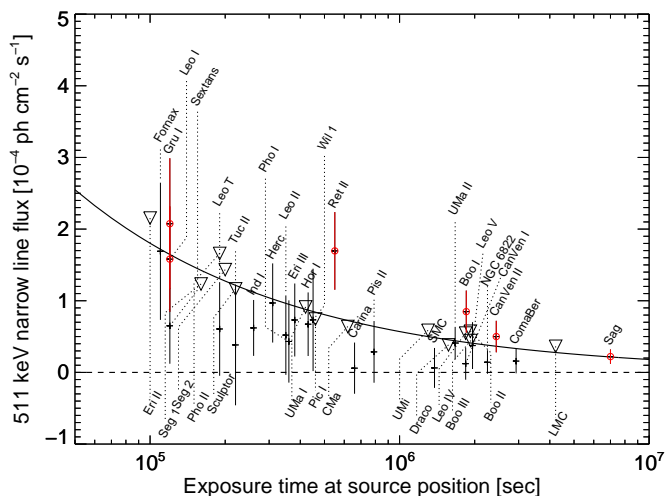}
 \caption{Derived fluxes (crosses) of each satellite galaxy against the exposure time at source position. If a line is not detected or appears negative, a $2\sigma$ upper limit is given (triangle). The solid line represents the $2\sigma$ sensitivity limit for a narrow line (instrumental resolution) seen with SPI at 511~keV. The (red) circles indicate sources for which the statistical significance is higher than $2\sigma$.}
  \label{fig:sat_stats}
\end{figure}

The summary of fit results for all 39 tested satellite positions is listed in Tab.~\ref{tab:dwarf_fits} and illustrated in Fig.~\ref{fig:sat_stats}. The exposure across the entire sky in this data set varies by a factor of 50 among the candidate sources, and the sensitivity changes accordingly. We empirically determine a $2\sigma$ narrow 511~keV line detection sensitivity of $5.7 \times 10^{-5} \times \sqrt{10^6 / T_{Exp} [\mathrm{Ms}]}~\mathrm{ph~cm^{-2}~s^{-1}}$ (solid line in Fig.~\ref{fig:sat_stats}). Among our sample of 39 candidate sources, 17 show weak indications of annihilation signals ($\geq 1\sigma$), independent of the exposure time. Six sources show a signal with more than $2\sigma$ (Leo I, Gru I, CVn II, Sag), and two sources more than $3\sigma$ (Boo I, Ret II) statistical significance above instrumental background. The values for Boo I may be over-/underestimated due to source confusion with Boo II. Statistically, one would expect about two $2\sigma$ sources out of a sample of 39 from fluctuations of the background. Since we see six sources at a significance of at least $2\sigma$ (two expected), and 17 sources at a significance of at least $1\sigma$ (13 expected), the 511~keV signals are not consistent with background fluctuations only. On the other hand, the individual 511~keV signals per source are of too low significance to single them out, and thus we will use the full population of possible sources for further analyses (see Sec.~\ref{sec:mtl}). Furthermore, we discuss the $3.1\sigma$ signal from Ret II in Sec.~\ref{sec:retII}, separately. 

\subsection{Stacked Analysis}\label{sec:stack}

Under the assumption that satellite galaxies share a common mass scale \citep{Strigari2008_dm_dsph}, we analyse the spectra in a constrained maximum likelihood fit to search for a DM-related 511~keV signal. For this, we determine one global scaling parameter to the set of sources, which are normalised to a common flux value and then re-scaled by their distances $D^{-2}$. We estimate the total $\gamma$-ray flux in the vicinity of 511~keV that reaches us from the positions of the Milky Way satellites (see Eq.(\ref{eq:stack})), and also avoid source confusion as above. In the stacked spectrum of the satellite galaxies at a canonical distance of 100~kpc, we do not find a significant excess and provide a $2\sigma$ upper limit of the flux of $1.4 \times 10^{-4}~\mathrm{ph~cm^{-2}~s^{-1}}$. This is based on ignoring DSGs closer than 25 kpc and DSGs in the direction of the galactic disk. Softening these restrictions by including all 39 DSGs changes this upper limit to $1.3 \times 10^{-4}~\mathrm{ph~cm^{-2}~s^{-1}}$. If the assumption of an identical DSG mass is discarded, Eq.~(\ref{eq:stack}) gets an additional factor $M_i^2$ where $M_i$ is the dynamical mass of the DSG. For a subset of galaxies with available J-factor and dynamical mass estimates (see Tab.~\ref{tab:dwarf_fits}), we derive an upper limit of $2.3 \times 10^{-4}~\mathrm{ph~cm^{-2}~s^{-1}}$. Under the same assumptions, with the requirement that DM annihilation explains the entire bulge signal \citep{Vincent2012_dm511,Evans2016_Jfactors}, the stacked dark matter signal would yield a 511 keV flux of $\sim 2 \times 10^{-6}~\mathrm{ph~cm^{-2}~s^{-1}}$.

\begin{table*}[!hb]
\caption{List of Milky Way satellites tested for 511~keV emission, ordered by distance. The measured line flux $F_{511}$ is given in $10^{-5}~\mathrm{ph~cm^{-2}~s^{-1}}$. $M_{Dyn}$ are the dynamical masses of the satellite in units of $10^6~\mathrm{\Msol}$, $M_V$ their absolute visual magnitude, and $d$ the distance in kpc. The significance of a possible line detection is given in units of sigma. $2\sigma$ or above detections are marked boldface. If a line is not present at all, a $2\sigma$ upper limit on the flux is given. The positions of the assumed centres of the satellites are given in galactic longitude $l$ and latitude $b$ in units of degrees. The effective exposure time at the position of the sources $T_{Exp}$ is given in Ms. $M_{Dyn}$, $M_V$, $d$, $l$, and $b$ are taken from the literature (references, last column). The distances have been chosen as the given mean value from the NASA/IPAC Extragalactic Database (NED), if available.}              
\label{tab:dwarf_fits}      
\centering                                      
\begin{tabular}{l | r r r r r r r r r}          

\hline\hline                        
Name                       & $d$            & $F_{511}$            & $M_{Dyn}$       & $M_V$            & $\sigma$       & $l$               & $b$               & $T_{Exp}$       & Ref.\\
\hline                                   
Canis Major\tablefootmark{b}                &   $9$          & $<4.1$               & $>49$           & $-14.4$          & $-$            & $239.99$          &  $-8.00$          & $0.62$          & (1),(16),(17) \\  
Segue 1\tablefootmark{b}                    &  $23$          & $<12.4$              & $0.26$          & $-1.5$           & $-$            & $220.48$          &  $50.43$          & $0.16$          & (1),(12),(60),(61),(62),(63) \\
\textbf{Sagittarius Dwarf} &  $\textbf{28}$ & $\textbf{2.2(1.0)}$  & $\textbf{190}$  & $\textbf{-13.4}$ & $\textbf{2.3}$ & $\textbf{5.57}$   & $\textbf{-14.17}$ & $\textbf{7.00}$ & (1),(44),(45),(46) \\  
\textbf{Reticulum II\tablefootmark{c}}      &  $\textbf{30}$ & $\textbf{17.0(5.4)}$ & $\textbf{0.24}$ & $\textbf{-2.7}$  & $\textbf{3.1}$ & $\textbf{266.30}$ & $\textbf{-49.73}$ & $\textbf{0.55}$ & (22),(23),(27),(42),(43) \\  
Ursa Major II\tablefootmark{c}              &  $34$          & $4.1(2.3)$           & $3.9$           & $-4.2$           & $1.9$          & $152.46$          &  $37.44$          & $1.67$          & (1),(57),(58),(59) \\  
Segue 2\tablefootmark{c}                    &  $35$          & $<14.4$              & $0.23$          & $-2.5$           & $-$            & $149.43$          & $-38.14$          & $0.20$          & (1),(48) \\  
Willman 1\tablefootmark{c}                  &  $42$          & $7.3(7.1)$           & $0.39$          & $-2.7$           & $1.0$          & $158.58$          &  $56.78$          & $0.45$          & (1),(62),(64),(65) \\
Coma Berenices\tablefootmark{c}             &  $44$          & $1.6(1.7)$           & $0.94$          & $-4.1$           & $1.0$          & $241.89$          &  $83.61$          & $2.93$          & (1),(6),(12),(18) \\  
Bo\"otes III               &  $48$          & $<4.4$               & $>0.017$        & $-5.8$           & $-$            &  $35.41$          &  $75.35$          & $1.93$          & (1),(8),(9),(10) \\  
Bo\"otes II\tablefootmark{a}                &  $49$          & $<5.8$               & $3.3$           & $-2.7$           & $-$            & $353.69$          &  $68.87$          & $1.92$          & (1),(5),(6),(7) \\  
Large Magellanic Cloud     &  $50$          & $<3.6$               & $>1500$         & $-18.1$          & $-$            & $280.47$          & $-32.89$          & $4.22$          & (1),(37),(38) \\  
Tucana II\tablefootmark{c}                  &  $57$          & $3.8(8.4)	$          & $N/A$           & $-3.8$           & $0.5$          & $328.08$          & $-52.32$          & $0.22$          & (22),(23) \\  
Small Magellanic Cloud     &  $61$          & $0.6(2.8)$           & $1400$          & $-16.8$          & $0.2$          & $302.80$          & $-44.30$          & $1.38$          & (1),(37),(52),(53) \\  
\textbf{Bo\"otes I\tablefootmark{a}\tablefootmark{c}}        &  $\textbf{62}$ & $\textbf{8.5(2.9)}$  & $\textbf{0.81}$ & $\textbf{-6.3}$  & $\textbf{3.0}$ & $\textbf{358.08}$ & $\textbf{69.62}$  & $\textbf{1.85}$ & (1),(2),(3),(4)  \\  
Ursa Minor\tablefootmark{c}                 &  $73$          & $<5.8$               & $9.5$           & $-8.8$           & $-$            & $104.97$          &  $44.80$          & $1.30$          & (1),(29) \\  
Horologium I\tablefootmark{c}               &  $79$          & $6.7(4.4)$           & $0.55$          & $-3.4$           & $1.6$          & $271.39$          & $-54.73$          & $0.43$          & (22),(23),(27) \\  
Draco\tablefootmark{c}                      &  $82$          & $<3.8$               & $11$            & $-8.8$           & $-$            & $86.37$           &  $34.72$          & $1.57$          & (1),(19),(20),(21) \\  
Phoenix II                 &  $83$          & $<16.6$              & $N/A$           & $-2.8$           & $-$            & $323.68$          & $-59.75$          & $0.19$          & (22),(23) \\  
Sculptor\tablefootmark{c}                   &  $83$          & $<11.6$              & $14$            & $-11.1$          & $-$            & $287.54$          & $-83.16$          & $0.22$          & (1),(47) \\  
Sextans\tablefootmark{c}                    &  $85$          & $6.5(5.3)$           & $10.6$          & $-9.3$           & $1.2$          & $243.50$          &  $42.27$          & $0.12$          & (1),(49),(50),(51) \\  
Eridanus III               &  $87$          & $7.3(5.1)$           & $N/A$           & $-2.0$           & $1.5$          & $274.95$          & $-59.60$          & $0.38$          & (22),(23) \\  
Indus I                    & $100$          & $6.2(3.9)$           & $N/A$           & $-3.5$           & $1.6$          & $347.15$          & $-42.07$          & $0.26$          & (23),(23) \\  
Ursa Major I\tablefootmark{c}               & $101$          & $<9.2$               & $11$            & $-5.5$           & $-$            & $159.43$          &  $54.41$          & $0.42$          & (1),(6),(54),(55),(56) \\  
Carina\tablefootmark{c}                     & $103$          & $0.6(3.6)$           & $6.3$           & $-9.1$           & $0.2$          & $260.11$          & $-22.22$          & $0.66$          & (1),(14),(15) \\  
Pictoris I                 & $114$          & $<7.4$               & $N/A$           & $-3.1$           & $-$            & $257.29$          & $-40.64$          & $0.46$          & (22),(23) \\  
\textbf{Grus I\tablefootmark{c}}            & $\textbf{120}$ & $\textbf{20.8(9.1)}$ & $\textbf{N/A}$  & $\textbf{-3.4}$  & $\textbf{2.3}$ & $\textbf{338.68}$ & $\textbf{-58.25}$ & $\textbf{0.12}$ & (22),(23) \\  
Hercules                   & $136$          & $9.7(5.5)$           & $2.6$           & $-6.6$           & $1.8$          &  $28.73$          &  $36.87$          & $0.31$          & (1),(6),(12),(26) \\  
Fornax\tablefootmark{c}                     & $139$          & $16.9(9.6)$          & $56$            & $-13.4$          & $1.8$          & $237.10$          & $-65.65$          & $0.11$          & (1),(24),(25) \\  
\textbf{Canes Venatici II\tablefootmark{c}} & $\textbf{153}$ & $\textbf{5.0(2.2)}$  & $\textbf{0.91}$ & $\textbf{-4.9}$  & $\textbf{2.3}$ & $\textbf{113.58}$ & $\textbf{82.70}$  & $\textbf{2.44}$ & (1),(6),(12),(13) \\  
Leo IV\tablefootmark{c}                     & $155$          & $<5.4$               & $1.3$           & $-5.8$           & $-$            & $265.44$          &  $56.51$          & $1.84$          & (1),(6),(12),(13) \\  
Pisces II\tablefootmark{c}                  & $182$  			  & $2.9(4.3)$           & $>0.0086$       & $-5.0$           & $0.7$          &  $79.21$          & $-47.11$          & $0.79$          & (1),(39),(40),(41) \\  
Leo V\tablefootmark{c}                      & $186$          & $3.7(3.3)$           & $1.1$           & $-5.2$           & $1.1$          & $261.86$          &  $58.54$          & $1.96$          & (1),(35),(36) \\  
Canes Venatici I\tablefootmark{c}           & $216$          & $1.2(2.2)$           & $19$            & $-8.6$           & $0.6$          &  $74.31$          &  $79.82$          & $1.84$          & (1),(6),(11) \\  
Leo II\tablefootmark{c}                     & $218$          & $5.0(5.5)$           & $4.6$           & $-9.8$           & $0.9$          & $220.17$          &  $67.23$          & $0.35$          & (1),(31),(32) \\  
\textbf{Leo I\tablefootmark{c}}             & $\textbf{246}$ & $\textbf{15.8(7.4)}$ & $\textbf{12}$   & $\textbf{-12}$   & $\textbf{2.2}$ & $\textbf{225.99}$ &  $\textbf{49.11}$ & $\textbf{0.12}$ & (1),(28),(29),(30) \\  
Eridanus II                & $380$          & $<21.6$              & $N/A$           & $-6.6$           & $-$            & $249.78$          & $-51.65$          & $0.10$          & (22),(23) \\  
Leo T\tablefootmark{c}                      & $412$          & $6.1(6.5)$           & $3.9$           & $-8.0$           & $1.0$          & $214.85$          &  $43.66$          & $0.19$          & (1),(33),(34) \\
Phoenix I                  & $418$          & $4.3(5.7)$           & $9.7$           & $-9.9$           & $0.8$          & $272.16$          & $-68.95$          & $0.36$          & (1),(66),(67),(68),(69) \\
NGC 6822                   & $498$          & $1.4(1.6)$           & $3500$          & $-15.2$          & $0.9$          &  $25.34$          & $-18.40$          & $2.25$          & (1),(29),(69),(70),(71),(72) \\

\hline                                             
\end{tabular}
\tablefoot{(1) \citep{McConnachie2012_dsph}, (2) \citep{Belokurov2006_dsph_BooI}, (3) \citep{Fellhauer2008_dsph_BooI}, (4) \citep{DallOra2006_dsph_BooI}, (5) \citep{Walsh2007_dsph_BooII}, (6) \citep{Grcevich2009_dsph_BooII}, (7) \citep{Walsh2008_dsph_BooII}, (8) \citep{Grillmair2009_dsph_BooIII}, (9) \citep{Carlin2009_dsph_BooIII}, (10) \citep{Correnti2009_dsph_BooIII}, (11) \citep{Zucker2006_dsph_CanVenI}, (12) \citep{Belokurov2007_dsph_CanVenII}, (13) \citep{Okamoto2012_dsph_CanVenII}, (14) \citep{Kraan-Korteweg1979_dsph_Car}, (15) \citep{Mateo1998_dsph_Car}, (16) \citep{Martin2004_dsph_CMa}, (17) \citep{Martin2005_dsph_CMa}, (18) \citep{Musella2009_dsph_CBer}, (19) \citep{Cotton1999_dsph_Dra}, (20) \citep{Falco1999_dsph_Dra}, (21) \citep{Tyler2002_dsph_Dra}, (22) \citep{Koposov2015_dsph}, (23) \citep{Bechtol2015_dsph}, (24) \citep{Piatek2007_dsph_For}, (25) \citep{Poretti2008_dsph_For}, (26) \citep{Musella2012_dsph_Her}, (27) \citep{Koposov2015_dsph_HorI}, (28) \citep{Whiting2007_dsph_LeoI}, (29) \citep{Young2000_dsph_LeoI}, (30) \citep{Caputo1999_dsph_LeoI}, (31) \citep{Coleman2007_dsph_LeoII}, (32) \citep{Gullieuszik2008_dsph_LeoII}, (33) \citep{Irwin2007_dsph_LeoT}, (34) \citep{Clementini2012_dsph_LeoT}, (35) \citep{Belokurov2008_dsph_LeoV}, (36) \citep{deJong2010_dsph_LeoV}, (37) \citep{Richter1987_dsph_LMC}, (38) \citep{Feast1987_dsph_LMC}, (39) \citep{Belokurov2010_dsph_PisII}, (40) \citep{Kirby2015_dsph_PisII}, (41) \citep{Sand2012_dsph_PisII}, (42) \citep{Simon2015_dsph_RetII}, (43) \citep{Walker2015_dsph_RetII}, (44) \citep{Majewski2003_dsph_Sag}, (45) \citep{Ibata1994_dsph_Sag}, (46) \citep{Monaco2004_dsph_Sag}, (47) \citep{Queloz1995_dsph_Scu}, (48) \citep{Belokurov2009_dsph_Seg2}, (49) \citep{Irwin1990_dsph_Sex}, (50) \citep{Battaglia2011_dsph_Sex}, (51) \citep{Lee2003_dsph_Sex}, (52) \citep{Matsunaga2011_dsph_SMC}, (53) \citep{Bekki2009_dsph_SMC}, (54) \citep{Willman2005_dsph_UMaI}, (55) \citep{Kleyna2005_dsph_UMaI}, (56) \citep{Okamoto2008_dsph_UMaI}, (57) \citep{Penarrubia2006_dsph_UMaII}, (58) \citep{Fellhauer2007_dsph_UMaII}, (59) \citep{DallOra2012_dsph_UMaII}, (60) \citep{Norris2010_dsph_Seg1}, (61) \citep{deJong2008_dsph_Seg1}, (62) \citep{Martin2008_dsph_Seg1}, (63) \citep{Simon2011_dsph_Seg1}, (64) \citep{Willman2005_dsph_Wil1}, (65) \citep{Willman2011_dsph_Wil1}, (66) \citep{Cote1997_dsph_PhoI}, (67) \citep{Zaggia2011_dsph_PhoI}, (68) \citep{Gallart2001_dsph_PhoI}, (69) \citep{Mateo1998_dsphs}, (70) \citep{Koribalski2004_dsph_NGC6822}, (71) \citep{Rogstad1967_dsph_NGC6822}, (72) \citep{Veljanoski2015_dsph_NGC6822}
\tablefoottext{a}{The values for Boo I may be over- or underestimated due to source confusion with Boo II, being not separated by at least one PSF. Likewise, the value for Boo II may be wrong, too.}
\tablefoottext{b}{For the stacking analysis, these galaxies have been ignored to validate the flux limit.}
\tablefoottext{c}{These galaxies have been included in the mass- and distance-weighted stacking analysis due to available dynamical mass and J-factor estimates, see Sect.~\ref{sec:stack}.}
}
\end{table*}

\section{Discussion}

\subsection{Mass-to-Light-Ratios}\label{sec:mtl}

\begin{figure}[!ht]
  \centering
   \includegraphics[width=\linewidth,clip]{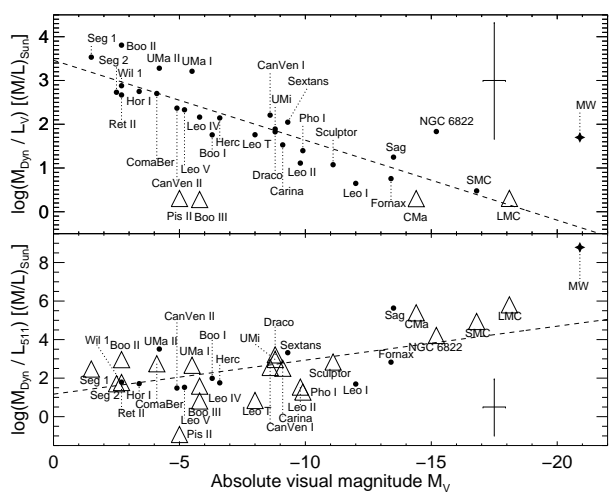}
 \caption{Mass-to-luminosity ratio in units of solar masses per solar luminosity as a function of absolute visual magnitude, $M_V$. Top panel shows the dynamical mass over the absolute V-band magnitude as already described by \citet{Mateo1998_dsphs}, \citet{Strigari2008_dm_dsph}, or \citet{McConnachie2012_dsph}. Towards fainter satellite galaxies, $\Upsilon_V$ increases, which is generally interpreted as indirect evidence for dark matter (see text for details). Bottom panel shows the ratio of the dynamical mass and the 511~keV luminosity over absolute visual magnitude. The trend is reversed when plotting $\Upsilon_{511}$ versus $M_V$, in contradiction with what is expected for a dark matter origin. Typical error bars are shown; $2\sigma$ lower limits are shown by triangles. For comparison, $\Upsilon_V$ and $\Upsilon_{511}$ for the Milky Way are shown with a star symbol in each panel.}
  \label{fig:mtl}
\end{figure}

The mass-to-light-ratio $\Upsilon_V = M_{Dyn} / L_V$ has been found to be a good indicator for DM which is believed to dominate the mass content in DSGs \citep{Mateo1998_dsphs,Strigari2008_dm_dsph,McConnachie2012_dsph}. In the top-panel of Fig.~\ref{fig:mtl}, the mass-to-light ratio within the half-light radius (see references in Tab.~\ref{tab:dwarf_fits}) against the absolute V-band magnitude from available literature data is shown. For Pis II, Boo III, CMa, and the LMC, no dynamical mass estimate is available and we used the stellar masses as lower limits for the dynamical masses. As already shown by several studies~\citep{Mateo1998_dsphs,Strigari2008_dm_dsph,McConnachie2012_dsph}, the mass-to-light-ratio shows a negative correlation with the brightness of the objects. This is counter-intuitive as naturally one would expect a nearly constant mass-to-light-ratio in the absence of dark matter, no matter how faint a galaxy is. The stellar-mass-to-light-ratio $\Upsilon_V^* = M^* / L_V$ indeed shows a value of \about 1.0 across the magnitude scale. But as the galaxies become fainter, $\Upsilon_V$ rises, which indicates an unseen mass, generally interpreted as DM sub-halos. We note that also the ultra-faint dwarf galaxies (data available for Hor I and Ret II), recently detected by \citet{Koposov2015_dsph}, nicely fit into this correlation. 

Any tracer that would make DM "visible", e.g. by measuring its annihilation products, should show a similar trend. We therefore define a mass-to-positron-annihilation-luminosity-ratio, $\Upsilon_{511} = M_{Dyn} / L_{511}$, and calculate these values for our sample. In the bottom panel of Fig.~\ref{fig:mtl}, we show $\Upsilon_{511}$ for the galaxies whose flux estimates deviate from zero (at the $1\sigma$ level). For all other galaxies for which data are available, we give $2\sigma$ lower limits. Apparently, and although the data have large uncertainties, the correlation is opposite to $\Upsilon_V$. The reversed trend for $\Upsilon_{511}$ versus $M_V$ is in contradiction with what is expected for a DM origin. This could have several causes: 
\begin{enumerate}
	\item The correlation is based on the high ratio derived from Sag; by neglecting this value, the rank correlation coefficient reduces from $-0.35$ to $-0.14$, but is still far from the positive correlation in the top panel. Using only signals with more than $2\sigma$ does also yield the same correlation.
	\item For the visually fainter galaxies (e.g. Ret II, Hor I) seen in 511~keV, the dynamical mass estimates are 2-3 orders of magnitude lower than for the bright galaxies (e.g. Sag, For) which automatically distorts the correlation in this direction, if the signals are not significant or strong.
	\item It is probably not the dynamical mass which drives the apparent correlation: As the correlation of $\Upsilon_V^*$ versus $M_V$ is completely gone, the respective correlation between $\Upsilon_{511}^*$ and $M_V$ is still there. Stars and their surrounding environments are a favoured explanation for any present 511~keV emission (see discussion about Ret II below), though the electron number density in DSGs is a crucial but uncertain factor in theoretical estimations of the annihilation rate.
\end{enumerate}

\subsection{Dark Matter Origin}

The pronounced spatial peak of the 511~keV signal in the galactic centre has been confirmed and strengthened by recent results (Paper I), reviving the possibility of a DM origin. If \pos s do not travel far from the source and rather find free or bound \ele s to annihilate with \citep{Guessoum1991_511ISM,Guessoum2005_511,Jean2009_511ISM,Alexis2014_511ISM}, the morphology would match the square of a host galaxy's DM density profile \citep[e.g.][]{Burkert1995_dm,Navarro1996_dm,Merritt2006_dm}. Interestingly, the peak of such a profile seen in 511~keV has been determined to be around $(l/b)=(-1.25/-0.25)^\circ$ \citep{Kuhlen2013_dm,Skinner2014_511}. \citet{Vincent2012_dm511} found that Einasto profiles also fit the data well, assuming a DM halo centred on the galactic centre position. Upper limits on the 1-2 MeV $\gamma$-ray continuum \citep{Boehm2004_dm,Boehm2008_dm,Beacom2006_511} limit the DM particle mass to $m_{DM} \lesssim 7~\mathrm{MeV~c^{-2}}$. These studies also conclude that the morphology of the signal precludes a decay-induced signal.

In the case of the DSGs, the signal would be seen by SPI as a point-like source, and the 511~keV flux, $F_{511}$, would follow $F_{511} = \frac{1}{4 \pi} \frac{\langle \sigma v \rangle}{m_{DM}^2}J$, assuming negligible positronium formation in the dwarfs, where $\langle \sigma v \rangle$ is the thermally averaged cross section, $m_{DM}$ is the DM particle mass, and $J$ the J-factor, see Eq.~(\ref{eq:jfactor}). \citet{Hooper2004_dm} estimated that if the whole 511~keV emission in the bulge of the MW was due to the annihilation of light DM particles into \ele\pos-pairs, an observable 511~keV emission form the direction of Sag would be about 3-6 times smaller than in the MW bulge. In our analysis, this ratio is $42\pm19$, ruling out this hypothesis by $\sim2\sigma$, though it is worth pointing out that the flux ratio between the GCS and Sag is $3.5\pm2.1$. If the Sag signal is entirely due to DM, this would indicate a DM contribution to the galactic signal of $\sim 3\%$. However, more recent fits to the bulge emission require a DM annihilation cross section that is a factor of 5 \citep{Ascasibar2006_511dm} to 10 \citep{Vincent2012_dm511} times smaller. The updated J-factor for Draco \citep{Ackermann2014_FermiLAT_dsphs,Evans2016_Jfactors} is furthermore $\sim 5$ times smaller than what was used by \citet{Hooper2004_dm}. This may also apply to Sag, although its morphological structure is more complex due to tidal stripping. Overall, this means that our measurement of the Sag flux does little to constrain the galactic centre signal.

Based on available J-factors \citep{Evans2016_Jfactors}, and assuming in-situ positron annihilation and negligible positronium formation, the strongest constraint we obtain on a DM origin comes from Ursa Major II, due to its large J-factor.  At $2\sigma$ confidence level, we derive
\begin{equation}
\langle \sigma v \rangle < 5.6 \times 10^{-28}\left(\frac{m_{DM}}{\mathrm{MeV}}\right)^2 \mathrm{cm}^3~\mathrm{s}^{-1}\mathrm{.}
\label{eq:ulumaii}
\end{equation}
This constraint is still two order of magnitude above the cross section required to explain the entire MW bulge signal, and could be weakened even further if the density of interstellar gas is too low for \pos s to efficiently find partners to annihilate with.

\subsection{Reticulum II}\label{sec:retII}

The ultra-faint dwarf galaxy Ret II \citep{Koposov2015_dsph,Simon2015_dsph_RetII} is found with a significance of 3.1 $\sigma$. This is tantalising evidence for a bright source of positrons in Ret II, and among the other DSGs, Ret II might be special from the perspective of two different, maybe unrelated, measurements:

\citet{Ji2016_RetII} measured strong enhancements of neutron-capture elements in stars of Ret II, and interpret this as the result of nucleosynthesis of heavy elements from a single enrichment event only, which then would have to be a neutron star merger. The same enrichment event could be a positron source, e.g. through evolving into an accreting black hole system, or else the existence of such neutron star binary also makes plausible the existence of a microquasar, producing \pos s in flaring states. On the other hand, there are suggestions for a star formation connection: \citet{Geringer-Sameth2015_RetII} reported a 2-10~GeV $\gamma$-rays with Fermi/LAT at 2.3 to 3.7$\sigma$ significance, and such $\gamma$-rays have been associated with star formation through cosmic-ray/gas interactions \citep{Abdo2010_GeV_M82,Ackermann2012_GeV_SF,Ackermann2016_GeV_LMC}. The effects of star formation are a non-negligible prerequisite for the 511~keV emission in the MW, as $\beta^+$-unstable radioactive nuclei are produced mainly in massive stars and their supernovae, and definitely contribute to the \pos-content in our Galaxy \citep[see e.g.][]{Diehl2006_26Al,Prantzos2011_511,Churazov2011_511,Alexis2014_511ISM}. 

At a distance of 30 kpc, Ret II shows a present-day positron annihilation rate (assuming a positronium fraction of 1.0) of $(3.7\pm1.2) \times 10^{43}~\mathrm{e^+~s^{-1}}$. This value is at least as high as the one for the entire MW ($(3.5-6.0) \times 10^{43}~\mathrm{e^+~s^{-1}}$, see Paper I), and would support either the neutron star merger hypothesis of \citet{Ji2016_RetII} or the star formation picture of \citet{Geringer-Sameth2015_RetII}. Either case may have produced a huge number of \pos s whose gradual, ongoing annihilation we now see in the ISM of Ret II. 

Although the GeV excess in Ret II may also be attributed to DM particle annihilation, the Fermi/LAT data itself does not favour one or the other annihilation channel, because of the large uncertainty in the DM content (J-factor) of Ret II \citep{Geringer-Sameth2015_RetII}. Furthermore, Ret II and the LMC are the only DSGs that show a high-energy excess, disfavouring a DM explanation of the signal, as otherwise more DSGs should have been detected \citep{Ackermann2014_FermiLAT_dsphs}. Using the J-factors from \citet{Evans2016_Jfactors}, a DM-only interpretation of the 511~keV signal from Ret II yields a cross section that would require a galactic bulge signal $\sim 100$ times larger than observed. Indeed, this would indicate that at most $\sim 1\%$ of Ret II's signal is due to DM annihilation.

\section{Conclusion}
\label{sec:fazit}

We reported a search for 511~keV electron-positron annihilation emission from the satellite galaxies of the Milky Way within 500 kpc. Out of 39 tested sources, we find a signal from only one galaxy, Reticulum II, with a significance of $3.1\sigma$. The results for all other satellite galaxies are not in contradiction although not entirely consistent with statistical fluctuations of background. A combined (stacking) analysis of the satellite galaxies, assuming they share a common dark matter mass scale \citep{Strigari2008_dm_dsph}, also does not yield a positive signal, and we provide a $2\sigma$ upper limit on the dark-matter related 511~keV line flux of $1.4 \times 10^{-4}~\mathrm{ph~cm^{-2}~s^{-1}}$. For a subset of galaxies with available masses and J-factors, we estimate a mass- and distance-weighted upper limit on the flux of $2.3 \times 10^{-4}~\mathrm{ph~cm^{-2}~s^{-1}}$ (see included galaxies in Tab.~\ref{tab:dwarf_fits}). Even when we tentatively accept all marginal signals, the measured fluxes do not scale with the distances to the satellite galaxies. Furthermore, the closest satellite galaxy in our sample, Canis Major, does not show any signal ($<4.1\times10^{-5}~\mathrm{ph~cm^{-2}~s^{-1}}$ at 2$\sigma$), though it might be influenced by extended emission from the galactic plane.

We have established a firm upper limit on the 511~keV emission from a dark matter origin, though more sensitivity will be required to test the dark matter hypothesis as the origin of the signal. The case of Reticulum II and the 511~keV signal from this galaxy cannot entirely be attributed to dark matter; other origins related to star formation or a single neutron star merger \citep{Geringer-Sameth2015_RetII,Ji2016_RetII}, are thus more plausible. Furthermore, we have used the constraints for the galactic centre 511~keV signal to show that the Reticulum II signal cannot be from dark matter alone. Understanding the signal of this dwarf galaxy may give clues about the true origin of the Milky Way bulge signal.

%
\begin{acknowledgements}
  This research was supported by the German DFG cluster of excellence 'Origin and Structure of the Universe'. The INTEGRAL/SPI project has been completed under the responsibility and leadership of CNES; we are grateful to ASI, CEA, CNES, DLR, ESA, INTA, NASA and OSTC for support of this ESA space science mission. MGHK acknowledges funding via the Australian Research Council grant DE130101399.
 \end{acknowledgements}
%
\bibliographystyle{aa}

%

\end{document}